\def\year{2020}\relax
\documentclass[letterpaper]{article} %
\usepackage{aaai20}  %
\usepackage{times}  %
\usepackage{helvet} %
\usepackage{courier}  %
\usepackage[hyphens]{url}  %
\usepackage{graphicx} %
\usepackage{verbatim}
\usepackage{subcaption}
\usepackage{graphicx}  %
\usepackage{booktabs}

\usepackage{graphicx}
\graphicspath{ {\string./figures/pdf/}, {\string./figures/rebuttal/} }
\usepackage{float}
\usepackage[show]{chato-notes}
\usepackage{amsmath,amssymb}
\usepackage{bbm}
\usepackage{epsfig}
\usepackage[linesnumbered,ruled,vlined]{algorithm2e} %
\SetKwInput{KwInput}{Input}                %
\SetKwInput{KwOutput}{Output}              %

\usepackage{algpseudocode}
\usepackage{url}
\usepackage{tikz}
\usepackage{multirow}
\usetikzlibrary{bayesnet}

\newcommand{\citet}[1]{\citeauthor{#1} \shortcite{#1}}
\newcommand{\citep}{\cite}

\newcommand{\spara}[1]{\smallskip\noindent{\bf #1}}

\newcommand{\PROD}{\textsf{PROD}}

\newcommand{\squishlist}{
 \begin{list}{$\bullet$}
  {  \setlength{\itemsep}{0pt}
     \setlength{\parsep}{3pt}
     \setlength{\topsep}{3pt}
     \setlength{\partopsep}{0pt}
     \setlength{\leftmargin}{2em}
     \setlength{\labelwidth}{1.5em}
     \setlength{\labelsep}{0.5em}
} }
\newcommand{\squishlisttight}{
 \begin{list}{$\bullet$}
  { \setlength{\itemsep}{0pt}
    \setlength{\parsep}{0pt}
    \setlength{\topsep}{0pt}
    \setlength{\partopsep}{0pt}
    \setlength{\leftmargin}{2em}
    \setlength{\labelwidth}{1.5em}
    \setlength{\labelsep}{0.5em}
} }

\newcommand{\squishdesc}{
 \begin{list}{}
  {  \setlength{\itemsep}{0pt}
     \setlength{\parsep}{3pt}
     \setlength{\topsep}{3pt}
     \setlength{\partopsep}{0pt}
     \setlength{\leftmargin}{1em}
     \setlength{\labelwidth}{1.5em}
     \setlength{\labelsep}{0.5em}
} }

\newcommand{\squishend}{
  \end{list}
}

\title{The Effect of People Recommenders on Echo Chambers and Polarization}

\author{Federico Cinus,\textsuperscript{1,2} Marco Minici,\textsuperscript{3,4} Corrado Monti,\textsuperscript{2} Francesco Bonchi\textsuperscript{2,5}
\\ \textsuperscript{1}Sapienza University, Rome, Italy; \textsuperscript{2}ISI Foundation, Turin, Italy; \textsuperscript{3}ICAR-CNR, Rende(CS), Italy
\\ \textsuperscript{4}University of Pisa, Italy; \textsuperscript{5}Eurecat, Barcelona, Spain
\\ federico.cinus@isi.it, marco.minici@icar.cnr.it, corrado.monti@isi.it, francesco.bonchi@isi.it
}

\begin{document}

\maketitle \sloppy

\begin{abstract}
The effects of social media on critical issues, such as polarization and misinformation, are under scrutiny due to the disruptive consequences that these phenomena can have on our societies. Among the algorithms routinely used by social media platforms, people-recommender systems are of special interest, as they directly contribute to the evolution of the social network structure, affecting the information and the opinions users are exposed to.
In this paper, we propose a  framework to assess the effect of people recommenders on the evolution of opinions. Our proposal is based on Monte Carlo simulations combining link recommendation and opinion-dynamics models. In order to control initial conditions, we define a random network model to generate graphs with opinions, with tunable amounts of modularity and homophily.
We join these elements into a methodology to study the effects of the recommender system on echo chambers and polarization.
We also show how to use our framework to measure, by means of simulations, the impact of different intervention strategies.

Our thorough experimentation shows that people recommenders can in fact lead to a significant increase in echo chambers.
However, this happens only if there is considerable initial homophily in the network.
Also, we find that if the network already contains echo chambers, the effect of the recommendation algorithm is negligible.
Such findings are robust to two very different opinion dynamics models, a bounded confidence model and an epistemological model.

\end{abstract}

\section{Introduction}
\label{sec:intro}

Social media have significantly transformed how the general public consumes information, to such an extent that many contemporary political events in the world have been connected to social media usage, from Arab Spring~\cite{khondkerRoleNewMedia2011} to Donald Trump election~\cite{enliTwitterArenaAuthentic2017}. Due to their disruptive potentiality, the algorithms adopted by social media platforms have been, rightfully, under scrutiny: in fact, such platforms are suspected of contributing to the polarization of opinions by means of the so-called \emph{``echo-chamber''} effect, due to which users tend to interact with like-minded individuals, reinforcing their own ideological viewpoint, and thus getting more and more polarized in the long run.
Among all types of algorithms potentially responsible for strengthening echo chambers,
\emph{people-recommender systems}  (e.g. ``People You May Know'' in Facebook or ``Who to Follow'' in Twitter)  are of special interest, as they directly contribute to the evolution of the social network structure, thus affecting the information and the opinions a user is exposed to. People recommenders mainly take advantage of two types of information to recommend who to follow: network structure (e.g., recommending friends of friends)  or content (e.g., recommending users with similar interests) \cite{BarbieriBM14}.
As such, one might expect that homophilic links are more likely to be recommended than heterophilic ones, contributing to the formation of echo chambers. As in a vicious loop, echo chambers might in turn make homophilic links more likely to be recommended in the future. To hold
algorithm designers and platforms accountable for the effects of their systems, we need to develop tools to measure
such effects.

Therefore, the main research question we tackle in this work is the following: \emph{how can we assess the effect of people recommenders on echo chambers and polarization?}

We tackle this research question by means of a principled approach that, by combining an
opinion dynamics model with a given people-recommender algorithm, is able to simulate the behavior of individuals changing their opinions as a consequence of their interactions with their neighborhood, within a social network that is continuously evolving.

\spara{Method.}
For what concerns opinion dynamics, we adopt two very different and complementary models:
the classic Bounded Confidence Model~\cite{deffuantMixingBeliefsInteracting2000b}, and the epistemological model by~\citet{bala1998learning}.
The former assumes opinions are equivalent; interaction happens by people close enough in their opinions (i.e., within their \emph{bounded confidence interval}) getting closer to each other.
The latter assumes that opinions are \emph{not} equivalent: one of the antipodes is a factual truth and the other is its negation. Each agent, while performing observations of the world, forms an opinion about a statement, that might be true or false.
During interactions, agents exchange the observations they made and update their opinions accordingly.
An example of the first model is a typical political debate (e.g., the Stay-Leave dichotomy during the Brexit campaign), while an example of the second model is the tension between a hoax (e.g., ``vaccines cause autism'') and the scientific evidence (e.g., ``vaccines do not cause autism").

As people recommenders, we adopt several state-of-the-art link predictors~\cite{liben2007link}
which are based on the network structure only, and one recommender which is biased by the opinions of the agents.

In order to assess the dependence of our results on initial conditions, we develop a random network model with opinions, extending the model by~\citet{lancichinettiBenchmarkGraphsTesting2008}.
Finally, we develop a series of metrics and techniques to examine whether
the presence of echo chambers and polarization in the network increases or decreases as a consequence of the recommender. Our approach, dubbed \PROD\ (as it enables to study the effects of  \textsf{P}eople \textsf{R}ecommenders on \textsf{O}pinion \textsf{D}ynamics), can also assess whether such consequences are statistically significant.

\spara{Findings.} Our main contribution is therefore the definition of a novel framework to assess the effect of any recommendation algorithm on echo chambers and polarization.
We do not place any assumption on the given recommender, which is seen as a black-box.
The results we can draw from our thorough experimentation are summarised as follows:

\begin{enumerate}
  \item People recommenders can strengthen echo chambers, as long as homophilic links are initially more present than heterophilic ones.

\item This effect becomes negligible (or even reversed) if the network is already segregated in polarized communities.

  \item The above findings are consistent between different opinion dynamics models and recommender algorithms.

  \item However, differences between recommenders can be observed: some recommenders dramatically contribute to polarization, while others do so only slightly and only under certain assumptions.
\end{enumerate}

Summing up, we show that people recommenders can amplify the intrinsic homophilic bias in a social network, contributing to the creation of echo chambers, where each node is surrounded by like-minded neighbors. Finally, we also showcase the usage of our framework to simulate the effects of three different intervention policies, aimed at mitigating echo-chambers and polarization effects. Our results show that a mitigation strategy which tries to recommend users to users with different opinions, might be effective in reducing the appearance of echo-chambers and polarization.

\section{Related Work}
\label{sec:related}

Despite the increasing attention received by the research community,  the role played by social media in opinion formation
remains an important open question. Many studies suggest that social media users tend to
access information from a narrow spectrum of opinions, forming \emph{``social bubbles''} of homogenous
individuals~\cite{nikolovMeasuringOnlineSocial2015,pariser11filter}, possibly as a
consequence of algorithmic filtering. In this perspective, since users within homogenous
bubbles tend to produce and consume more information that confirms their preconceptions, social media might turn into \emph{echo
chambers}~\cite{quattrociocchiEchoChambersFacebook2016}, reinforcing their beliefs.
However, this view is far from unanimous.
Some researchers consider that ideological segregation in social-media usage has been overestimated~\cite{barberaTweetingLeftRight2015,garrett09echo}.
\citet{bakshyExposureIdeologicallyDiverse2015a}, through an empirical analysis of Facebook data, suggests that algorithmic ranking produces, in fact, more exposition to diverse viewpoints.
\citet{fletcherArePeopleIncidentally2017} confirm, through surveys, that casual social media users get exposed to more points of view than similar people who do not use social media at all.
\citet{morales2021no} show that in the Reddit political community, users of opposite leaning are more likely to interact than users with similar leaning; also, they show that geographical echo chambers might actually play a larger role.
Others have noted that echo chambers observed online could be, in fact, originating offline~\cite{bastosGeographicEmbeddingOnline2018}, thus motivating the need to consider different possible pre-existing levels of homophily when studying the relationship between algorithms and polarization.
Finally, \citet{cinelli2021echo} found evidence of echo chambers on Facebook, but not on Reddit, showing that there exist differences between different social media platforms that might depend on different content recommendation mechanisms.

Our contribution in this debate lies in a principled approach to assess, through simulations, the effects of people-recommender algorithms on opinion dynamics.

Few recent studies address similar research questions.
\citet{sirbu2019algorithmic} show that algorithmic bias can slow down the convergence of an opinion dynamics model, and contribute to fragmentation, but they do not study the effect of people recommenders.

\citet{perra2019modelling} study the effect of different information filtering policies by designing a novel opinion dynamics model;
they find a strong effect when the information received by a user is biased towards their opinion, and that some network configurations can reinforce such effect.

\citet{FabbriBB020} study the effects of people recommender systems on the visibility of users in a network divided in two subgroups (e.g., males and females, or democrats and republicans), and with different levels of initial homophily.
They show that homophily plays a key role in the visibility given to different groups: when the minority is homophilic, there is a disparate visibility in favor of the minority class; when the minority is not homophilic, the dis-parate visibility is in favor of the majority class.

\citet{sasahara2021social} focus on user-driven rather than platform-induced behavior, showing that a small bias can lead to a segregated social network.

Finally, \citet{de2021modeling} recently proposed a novel opinion dynamics model that incorporates the interaction with an idealized recommender.
They found that the rise of echo chambers crucially depends on the role of the social network platform, expressed as a probability distribution of content being seen by others.
In this work, instead, we focus on \emph{existing} recommender algorithms, and study their effects on well-known opinion dynamics models, through their interactions with the network structure.

\section{Framework}
\label{sec:model}

\spara{Input.}
\PROD\ takes as main input a directed social graph  $G=(V, E, O)$ where an arc $(u, v) \in E$ indicates that $u$ follows $v$, while $O: V \rightarrow [0, 1]$ is a function assigning to each node in the network an opinion.
The other main input is a link recommender, that we can think as a function $\ell_G: V \rightarrow V \times [0, 1]$. Given a node $u \in V$, the function assigns a recommended node $v \in V$ such that $(u,v) \notin E$ and a probability $p_v \in [0, 1]$.
Such probability represents the strength of the recommendation given to $u$ to start following $v$ or, alternatively, the probability that $u$ will accept the recommendation and start following $v$, thus adding the new link $(u,v)$ to $E$.

 On top of these two main ingredients---a graph with opinions, and a recommender---\PROD\ adds a given opinion dynamics model (ODM).
  ODMs are expressed with an \emph{update rule}, which modifies the opinions of two nodes that interact.

  \spara{Framework overview.}
  Given this input, \PROD\ operates according to the pseudocode in Algorithm \ref{alg:prod}.

  For a number of time steps $T_{\max}$, nodes are visited sequentially in random order (lines 3-5).
  At each time step, every node interacts with $S$ other nodes, according to the update rule of the opinion dynamics model. There are two ways interactions can happen: through newly created links (lines 10-19) or through pre-existing links (lines 20-23). In the former case, the ODM and the link recommender work jointly, while the latter case is a typical step of an ODM.

 More in detail. In the first case, the recommender algorithm is asked for a node $v$ to recommend to $u$ (line 10).
  Together with the node $v$, the algorithm will also return the score $p_v\in[0, 1]$---the probability of the recommendation being accepted.
 If the recommendation $(u, v)$ is accepted, the interaction happens (according to the ODM), and the new link is added to the network.
  After doing this, we perform a \emph{rewiring}, in order to model the well-known concept of \emph{attention budget}: since time and attention are limited resources, users interact with a small set of neighbors~\cite{golder2007rhythms,huberman2008social}, even if nominally they declare a larger number of friends on social media.
 To keep this fact in consideration, whenever $u$ starts to follow another node $v$, we remove one of its old neighbors at random.
 This mechanism will also keep the overall density of the network stable, letting us attribute any observed effect to the recommender algorithm itself, rather than to a densification of the graph.

When a given number of total recommendations $R_{\max}$ is reached, the recommender stops and lets the ODM work alone, to let emerge any long-term effect of the final graph on the opinion distribution. We calibrate the internal parameter $\alpha$ such that these two regimes (the one with link recommender and ODM cooperating and the one purely ODM-based) last approximately $\frac{T_{\max}}{2}$ time steps each, to keep the simulations for different recommenders with the same amount of recommendations $R_{\max}$ and the same number interactions between nodes ($S \cdot |V| \cdot T_{\max}$).

In the rest of this section, we are going to discuss the two main ingredients of \PROD: the opinion dynamics model and the people-recommender algorithm.

\begin{algorithm}[t]
\DontPrintSemicolon

  \KwInput{Graph with opinions $G=(V, E, O)$, \\

  $\quad\quad\quad$people recommender $\ell_G: V \rightarrow V \times  [0, 1],$ \\
  $\textbf{Parameters:}$ interactions per time step $S \in \mathbb{N}^+$. \\
  $\quad\quad\quad$number of recommendations $R_{max} \in \mathbb{N}$, \\
  $\quad\quad\quad$number of time steps $T_{max} \in \mathbb{N}^+$.
  }
  $r, t \leftarrow 0$ \;
  $\alpha \leftarrow R_{max}/(\frac{T_{max}}{2} \cdot S \cdot |V|)$ \;
  \While{$t < T_{max}$}
   {
    $t\texttt{+} \texttt{+}$ \;
   	$\sigma \leftarrow$ random permutation of $V$ \;
   	\ForAll{$u \in \sigma$}{
   	    $s \leftarrow 0$ \;
   	    \While{$s < S$}{
           	\If{$\operatorname{Bernoulli}(\alpha)$}
            {
              $v, p_v \leftarrow \ell_G(u)$
            	
                \If{$\operatorname{Bernoulli}(p_v)$}
                 {
                     $w \leftarrow$ random node from $N_{out}(u)$ \;
                     $E \leftarrow E\setminus \{u, w\}$ \;
                     $E \leftarrow E\cup \{u, v\}$ \;
                     $\operatorname{UpdateRule}(O_u , O_v)$ \;
                     $s\texttt{+} \texttt{+}$ \;
                     $r\texttt{+} \texttt{+}$ \;
                     \If{$r = R_{max}$}{
                         $\alpha \leftarrow 0$ \;
                      }
                 }
            }
            \Else
            {
               $v \leftarrow$ random node from $N_{out}(u)$ \;
               $\operatorname{UpdateRule}(O_u , O_v)$ \;
               $s\texttt{+} \texttt{+}$ \;
               		
            }
        }
   	}
   }

\caption{PROD}
\label{alg:prod}
\end{algorithm}

\subsection{Opinion dynamics models}\label{sec:odm}

We consider two complementary models: the classic Bounded Confidence Model~\cite{deffuantMixingBeliefsInteracting2000b}, and the epistemological model by~\citet{bala1998learning}. As already discussed in Section \ref{sec:intro}
 these two models are very different and complementary. We next see them in details.

\spara{Bounded Confidence Model (BCM).}
 In BCM, interactions modify the opinions of the nodes only when they are within a confidence interval $\epsilon \in [0, 1]$ from each other. If they are, when $u$ interacts with $v$, $u$ moves closer to $v$'s opinion.
The strength of the interaction is represented by convergence parameter $\mu \in [0, 0.5]$.
The update rule of BCM is therefore defined by:
\begin{equation*}
    o^{new}_u =
\begin{cases}
    o_u + \mu \cdot (o_v - o_u) & \text{if } |o_u - o_v| < \epsilon \\
    o_u                        & \text{otherwise}
\end{cases}
\end{equation*}

\spara{Epistemological model.}
This model describes how a network of agents would form an opinion about a statement, that might be true or false. Consider for example a group of clinicians which use a drug X with a constant and known probability of success ($p_{succ}=0.5$). Suppose that a new drug Y, with an unknown probability of success, is introduced. Each clinician tries the new drug in some experiments, shares the results with his network, and then updates his belief on the new drug using all the data acquired.

As in the model introduced by \citet{bala1998learning}, we consider:
 (i) two Bernoulli random variables $\{x_0, x_1\}$, which represent the actions of the agents (e.g. using the known drug or experimenting with the new one);
 (ii) a ``true'' state of the world $\theta$, that defines their probability of success $\theta=[p_{succ}(x_0)=0.5, p_{succ}(x_1)=0.5+\epsilon]$.
 The parameter $\epsilon \in (0, 0.5]$ represents the gain of $x_1$ over $x_0$ in expectation, but it is completely unknown to the agents; in fact, agents need to form an opinion on whether $\epsilon > 0$ is true, or not.

 Each agent, in fact, has a (normalized) belief $o(\theta)$ on the state of the world $\theta$, which drives their next action.
 Two possible outcomes can arise: (i) the agent hypothesizes that $\epsilon$ is negative (i.e. the probability of success of the new drug is less than the previous), and they performs action $x_0$ for $n$ times, giving a fixed outcome of $\frac{n}{2}$; (ii) their hypothesis is that $\epsilon > 0$, which implies that $p_{succ}(x_1)>p_{succ}(x_0)$ and therefore they perform action $x_1$.
 In this case, the outcome of the $n$ experiments is a Binomial distribution with $p_{succ}(x_1)=0.5+\epsilon$.

 Finally, when agent $u$ interacts with $v$ over the arc $(u, v)$, they update their beliefs by considering the experiments performed by $v$ (if any), according to a Bayesian learning model:
\begin{equation}
o(\theta)^{new}_u = \dfrac{1}{1+\dfrac{1-o(\theta)}{o(\theta)}\left(\dfrac{0.5-\epsilon}{0.5+\epsilon}\right)^{2k_v-n}}
\label{eq:bayes}
\end{equation}
where $k_v$ is the number of successes obtained by $v$ in their $n$ experiments.

\subsection{People-recommender algorithms}
\label{sec:recommenders}
People recommenders are key services in social media and social networking platforms. Although early people recommenders were essentially performing link prediction using only the graph structure as input information~\cite{liben2007link}, modern algorithms often combine different sources of information through learning techniques: see~\cite{guy2018people} for a recent survey.
However, for sake of simplicity, we can consider two main types of information as input to the people recommenders: the network structure (e.g., recommending friends of friends),  content, activity and behavioural information (e.g., recommending users with similar interests).

Given that in our model we only have the graph structure and user opinions, we consider three scalable state-of-the-art link prediction methods that only use the network structure ---Jaccard Index, Personalized PageRank \cite{kumar2020link}, and SALSA ~\cite{wtf_twitter,guy2018people}--- and a method which relies on users opinions. All the methods we consider (presented in details next) accept as input a directed social graph.

\spara{Directed Jaccard index (DJI).}
Let us indicate with
$N^+(u)$ the out-neighbors set for the node $u$, and with $N^-(u)$ its in-neighbors set, on the given directed graph $G=(V, E)$.

Then, we define their \textit{Directed Jaccard Index} as
\begin{equation*}
DJI(u,v) = \frac{
  | N^+(u) \cap N^-(u) |
}{
  |N^+(u)| + |N^-(v)| - | N^+(u) \cap N^-(u) |
}
\end{equation*}

\spara{Personalized PageRank (PPR).}
Given the adjacency matrix $A$ associated to a directed graph $G=(V, E)$, where each entry $a_{ij}=1$ iff $(i,j) \in E$, and its column-stochastic version $A'$ (i.e. each entry is the probability to transition from $i$ to $j$), the \textit{global PageRank vector} $\mathbf{p}$ is the solution of a linear system \cite{page1999pagerank,gleich2015pagerank} and satisfies
 \begin{equation}
 \mathbf{p} = d \cdot A' \mathbf{p} + (1-d )\cdot \mathbf{r}
 \label{eq:pr}
 \end{equation}
where $d$ is called \textit{damping factor} and $1-d$ is the \textit{teleportation probability}, representing the probability that a random walker transits randomly to another node extracted with uniform distribution $\mathbf{r}$ defined over $V$. Setting $\mathbf{r}=\mathbbm{1}_{u}$, where $\mathbbm{1}_{u}$ is a $|V|$-dimensional vector with $1$ at the $u^{th}$ entry, reduces the solution of Eq.~\ref{eq:pr} to the \textit{Personalized PageRank (PPR)} for node $u$.

\spara{SALSA.}
In \cite{wtf_twitter}, the Who To Follow service technology architecture implemented by Twitter is depicted along with the algorithm used for the recommendations. The proposed model makes use of \textit{SALSA} algorithm proposed by \citet{salsa}.
\textit{SALSA} is a link-analysis algorithm, of the same family of the well-known \textit{HITS} \cite{hits}. The algorithm used in our work follows the schema described by the same authors of \cite{wtf_twitter} in a subsequent paper \cite{wtf_2015}, using a score-propagation strategy.
Given a user $u \in V$, \textit{SALSA} builds a bipartite graph $G_b = (V_h, V_a, E_b)$, where $V_h$ is the set of \emph{hubs}, i.e. the nodes with top-$k$ value of Personalized PageRank on node $u$, and $V_a$ is the set of \emph{authorities}, i.e. the nodes the hubs follows in the original graph $G$.
Each directed edge $(i,j) \in E$ becomes a directed edge $(i_h,j_a) \in E_b$.
 Let $M$ be the matrix with dimension $(|V_a|, |V_h|)$, where $M_{ij} = 1$ iff $(j_h, i_a) \in E_b$, zero otherwise. Let $M'$ be the matrix $M$ with each non-zero column divided by the sum of the column. Let $M^T$ be the transpose of the matrix $M$ with each non-zero row divided by the sum of the row. Let $d$ be the \textit{damping} column vector with $d_i = \hat{d}$ if $i = u$ and zero otherwise. Two equations are defined as follows:
 \begin{equation}
 \label{eq:salsa1}
 \mathbf{r} = M'  \mathbf{s}
 \end{equation}
 \begin{equation}
 \label{eq:salsa2}
 \mathbf{s} = d + (1 - \hat{d}) \cdot M^T \mathbf{r}
 \end{equation}
$\mathbf{r}$ and $\mathbf{s}$ are two vectors of size $|V_a|$ and $|V_h|$ which represent the relevance scores of authorities and similarity scores of hubs. The procedure initializes $\mathbf{s}$ with $s_i = 1$ iff $i = u$ and zero otherwise. Equations \ref{eq:salsa1} and \ref{eq:salsa2} are then recursively applied until the values converge.
  The list of recommendations is the set of authorities $V_a$ sorted by the values in $\mathbf{r}$.

 \spara{Opinion-biased algorithm (OBA).}
Besides these network-based classic recommenders, we also consider an idealized algorithm that recommends links by matching nodes with a similar opinion.
 Considering an opinion vector $\mathbf{o}$, the opinion distance between $u$ and $v$ is simply the difference $d_{uv}=|o_u - o_v|$. Therefore, as in~\cite{sirbu2019algorithmic}, we define their opinion-biased recommendation score as
 \begin{equation*}
 p_{uv} = \frac{d_{uv}^{-\gamma}}{\sum_{(u,v)\in E}d_{uv}^{-\gamma}}
\end{equation*}
where $\gamma$ tunes the importance of similarity: as $\gamma$ increases, the recommendation between similar nodes is more likely.

\spara{Normalization of recommenders.}
Finally, an important issue is how to make the effect of the recommenders comparable.
As stated before, each recommender's procedure derives different recommendation scores, expressing the strengths of their recommendations. Indeed, every recommendation algorithm generates a different list of ranked links for each node. Even if their scores are bounded, the scale and distribution of these scores may intrinsically depend on the nature of the recommender and its similarity index.

To ensure that all results are comparable among the different models, we implement a quantile transformation of the recommendation probabilities' distribution, which allows us to keep the monotonicity of the ranking scores and scale the distribution to a uniform one.
Specifically, we use a probability integral transform: given a random variable $X$ distributed according to the cumulative distribution function $F_X$ (CDF), then we define $Y=F_{X}(X)$, thus having a uniform distribution.
In practice, to empirically estimate $F_X$ for a given recommender, we draw a large sample of recommendation scores at the beginning of the simulation.
Then, each time a recommender is used in the process, the probability of recommendation is transformed accordingly.

\section{Evaluation Method}
\label{sec:framework}

In this section, we are going to discuss how to use \PROD\ in order to evaluate the effect of a given recommender algorithm on echo chambers and polarization.
The proposed evaluation methodology consists of the following steps:

\begin{enumerate}
  \item Consider a grid of parameters $\mu, \eta$, representing networks with different modularity and initial homophily (see details in Section~\ref{sec:network-model}).
  \item For each pair of parameters $\mu, \eta$ generate a sequence $\mathcal{G}_{\mu, \eta}$ of $K$ random graphs with opinions.
  \item On each graph $G \in \mathcal{G}_{\mu, \eta}$, run \PROD\ (Algorithm 1) without any recommender algorithm (i.e., with $R_{\max}=0$), obtaining a network $G'_0$.
  \item Then, on each graph $G \in \mathcal{G}_{\mu, \eta}$, run \PROD\ with the studied recommender algorithm $\ell$, obtaining $G'_{\ell}$.
  \item Considering a metric $m$ for echo chambers or polarization (see section~\ref{sec:metrics}), compute $m(G), m(G'_0), m(G'_{\ell})$.
  \item Assess the average difference $\Delta m$ between the two cases.
  Since the initial measure $m(G)$ is the same with and without recommender, we obtain
  \begin{equation}
    \label{eq:difference}
    \Delta m =
     \frac{1}{K}
     \sum_{G \in \mathcal{G}_{\mu, \eta}} m(G'_{\ell}) - m(G'_0).
  \end{equation}
\item Perform a \emph{Kolmogorov-Smirnov test} comparing the distribution of differences in the two cases, determining whether the observed effect is statistically significant.
\end{enumerate}

\noindent In the remainder of this section we are going to discuss the random network model we designed, which extends the model by~\citet{lancichinettiBenchmarkGraphsTesting2008} with opinions.
Then, we are going to discuss the two metrics we use: the \emph{neighbor correlation index} (\textsc{NCI}), and the \emph{random walk controversy score} (\textsc{RWC})~\cite{garimella2018quantifying}.

\subsection{Random network model}
\label{sec:network-model}

In order to study the effect of people recommenders in all possible configurations of homophily, we need a random network model with different characteristics.
First, it needs to have a tunable degree of echo chamber structure.
Second, it has to define opinions as well, coherently with the network structure.
Finally, it has to be realistic in its basic features with respect to real social networks.
For these reasons, we extended the LFR model proposed by~\citet{lancichinettiBenchmarkGraphsTesting2008}.
This model allows creating networks with a realistic degree distribution, following a power-law, and where each node belongs to a community.
In the original work, the authors show the realistic properties of this model.
Moreover, its community structure lets us easily define the opinions of each node.

The general assumption is that each node might follow the opinion of their community or might develop an opinion independently from their neighbors.
In this way, we decouple and distinguish two main components of the echo-chamber structure of the network:
its (initial) \emph{homophily}, describing how close on average is a node opinion from the opinions of their local community; and its \emph{modularity}, that describes how segregated are the different communities.
The former is introduced by our opinion-aware extension of the original model, while the latter is a parameter of the LFR model.
We indicate these parameters $\eta$ and $\mu$, respectively.

The procedure takes as input three parameters: the number of nodes in the graph $N$, the modularity $\mu$ and the initial homophily $\eta$, while the output consists of a directed graph $G$ and opinions $O \in [0,1]^N$ of the nodes.
The modularity parameter of the original model $\mu$ tunes the ratio of intra-community edges; increasing this value makes the network more split in communities.
From the original model, we obtain a set of edges $E$ and a partition of nodes into communities $c: V \rightarrow C$.
We note that in the LFR model the sizes of communities are realistic and follow a power-law distribution.
To generate opinions, we first pick an opinion for each community $k \in C$ by drawing $o_k \sim U(0,1)$.
Then, the opinion of a node $v$ is decided by a Bernoulli trial with probability $\eta$.
If the outcome is successful, the node assumes the opinion of its community, i.e. $o_v = o_{c(v)}$; otherwise, it draws a individual opinion $o_v \sim U(0,1)$.
Hence, higher values of $\eta$ increase the probability that a node shares the opinion of the other nodes belonging to the same community.

\begin{figure}[t!]
\centering
\includegraphics[width=\columnwidth]{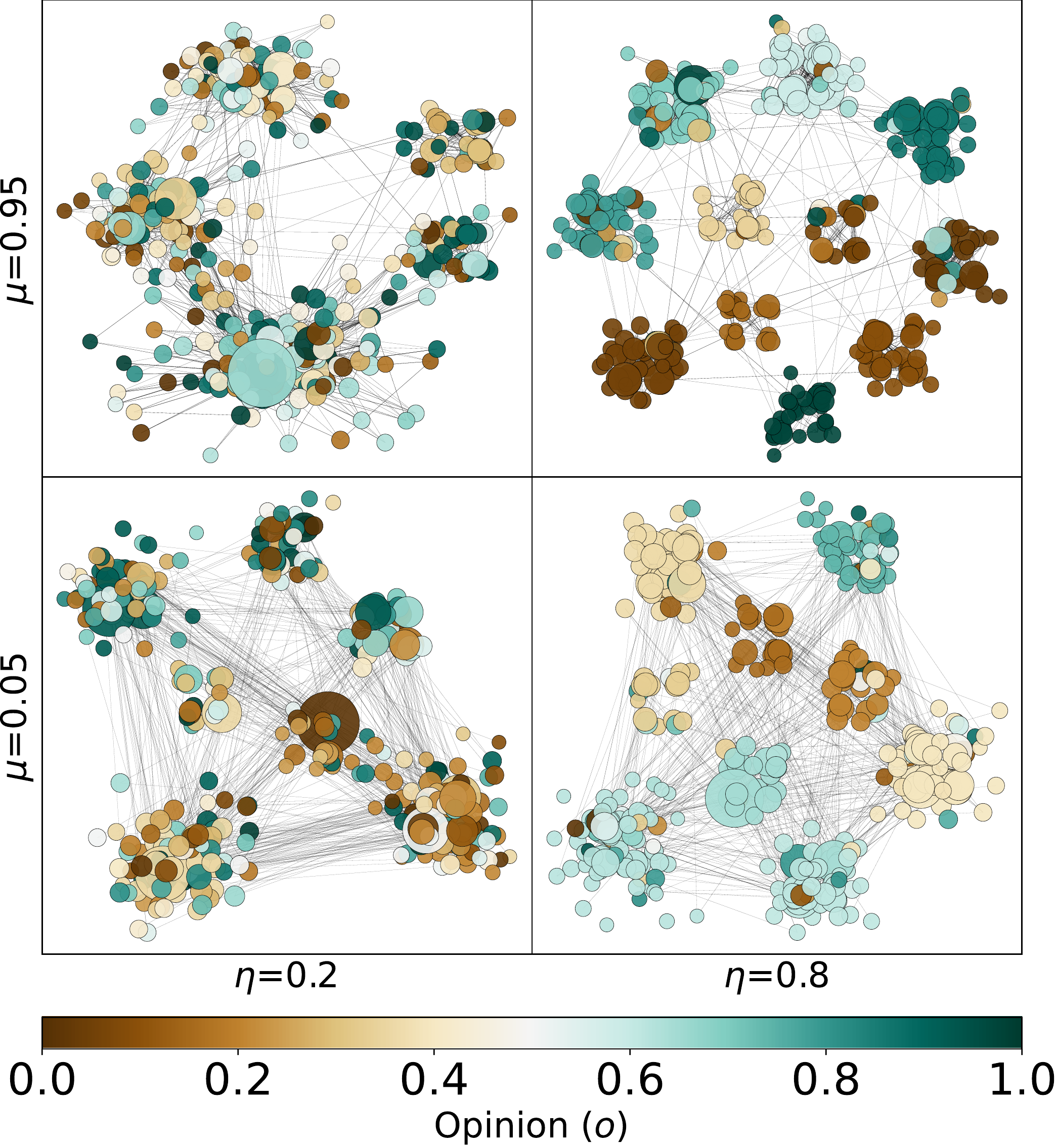}
\caption{Examples of graphs and opinions obtained with the random network model in four corner cases of the parameter space (homophily $\eta$ on the x-axis, modularity $\mu$ on the y-axis).
Colors correspond to the opinion of each node; the size is proportional to the degree of each node.
Each community produced by the model has an assigned position in the axis derived from the Fruchterman-Reingold force-directed algorithm~\cite{kobourov2012spring}. For each node, we draw random coordinates around the correspondent center of the assigned community. Here the density is constant in all graphs.}
\label{fig:random-graphs}
\end{figure}

With this procedure, we can generate random networks with opinions, by controlling both their initial homophily $\eta$ and their modularity $\mu$.
A network with high values for both will have a very polarized structure.
Instead, for instance, low values for $\eta$ and high values for $\mu$ will lead to graphs with well-separated communities, but plenty of variance inside each.
We show an example network for each of the four corner cases in Figure~\ref{fig:random-graphs}.

\begin{figure*}[t]
\centering
\begin{tabular}{cc}
 \hspace{-8mm}\includegraphics[width=.55\linewidth]{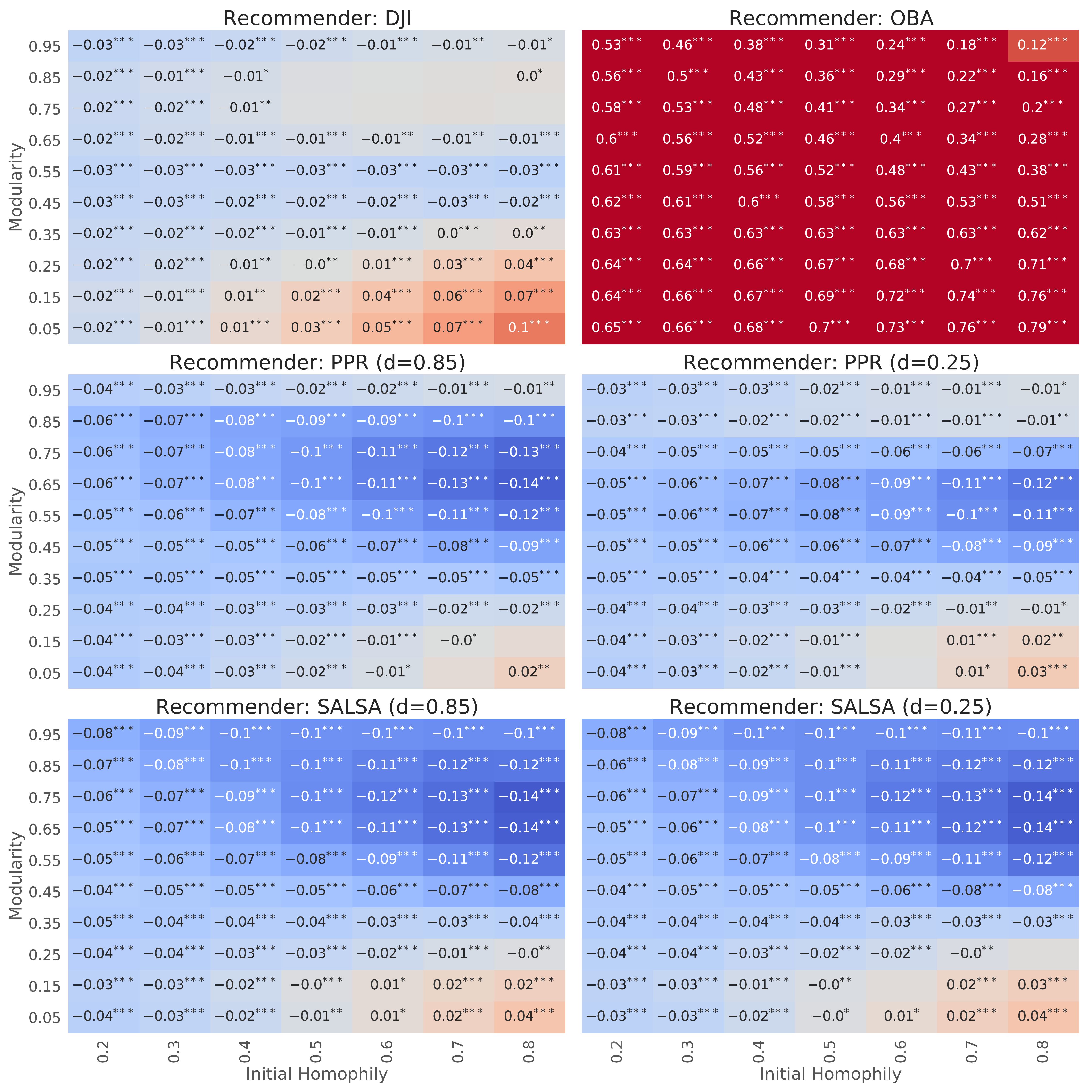}&
\hspace{-6mm} \includegraphics[width=.55\linewidth]{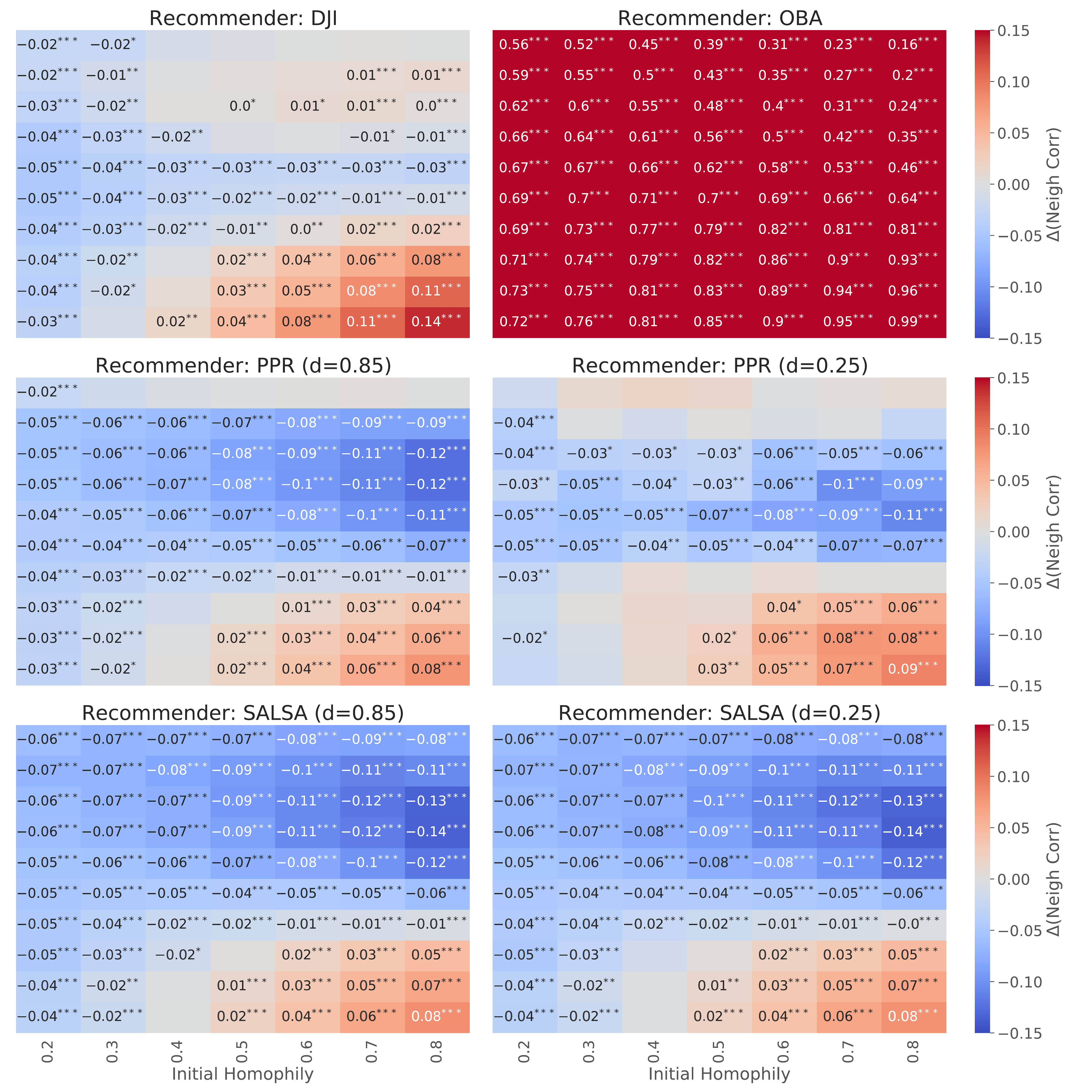}
\end{tabular}
\caption{\label{fig:nci}
$\Delta$NCI for the epistemological model (left half) and for the BCM model (right half). Each heatmap represents a given recommender and colors represent $\Delta$NCI
under different initial conditions (homophily $\eta$ on $x$-axis, modularity $\mu$ on $y$-axis). Numbers are shown when the difference is statistically significant ($p < 0.05$). Asterisks further represent levels of significance (** is $p < 0.01$, *** is $p < 0.001$).}
\end{figure*}

\subsection{Metrics}
\label{sec:metrics}
In order to measure the effect of the recommender systems in terms of echo chambers and polarization phenomena, we employ two global metrics defined over a graph where each node is labeled with an opinion. The first one, that we introduce, is Neighbor Correlation Index (NCI), which measures the similarity between each node and its neighbors, thus quantifying the echo-chamber effect.
 The second one, borrowed from the literature, is the Random Walk Controversy score (RWC), which measures the polarization of the network, in terms of random walks.

\spara{Neighbors Correlation Index (NCI).} We define NCI as the Pearson correlation $\rho(\mathbf{o}, \mathbf{o}^{N}_{u})$ between the opinion vector $\mathbf{o}$, and the average opinion of each node's neighbors, that is
\begin{equation*}
  \mathbf{o}^{N}_{u} = \frac{1}{N(u)} \sum_{v\in N(u)} o_v.
\end{equation*}
 Therefore, NCI is bounded between $[ -1, 1]$.
 The value $-1$ represents perfect anticorrelation: each node has exactly the opposite opinion of its neighbors.
 Instead, $1$ represents perfect correlation: each node has exactly the same opinion of all the nodes they follow---i.e., they are embedded in perfect echo chambers.

\spara{Random Walk Controversy score (RWC).}
In order to measure polarization, we employ the Random Walk Controversy score as defined by \citet{garimella2018quantifying}.
 This measure accounts for the probability that a random user could be exposed to authoritative content from the opposite side. Given two disjoint components of the graph $X$ and $Y$, the Random Walk Controversy score is
\begin{equation*}
    RWC = P_{XX}P_{YY} - P_{XY}P_{YX}
\end{equation*}
where $P_{ij}$ is the probability for a random walker that ends in partition $j$ to have started in partition $i$.
 In our case, we define $X$ (resp. $Y$) as the set of nodes $v$ with opinion $o_v < 0.5$ (resp. $o_v > 0.5$).
 This measure is not influenced by the size of the components and the total degree of the nodes in the two partitions, and it is bounded between $-1$ and $1$.
 High values correspond to a low probability of crossing the partitions w.r.t. staying in the same partition: therefore, it means that the two sides are very well separated, and thus that the network is polarized.
 Oppositely, values around $0$ in this metric reflect an equal probability to cross sides and to stay in the same one.
 In practice, we follow the suggested implementation~\cite{garimella2018quantifying} and use random walk with restart, forcing a restart when the random walk reaches a high-degree node (specifically, over the $95^{th}$ percentile of the degree distribution).

\smallskip

 An important distinction between the two metrics is that since RWC is based on random walks, this metric incorporates similarities between distant nodes.
 Instead, NCI captures local dynamics, by measuring how much each node's opinion is echoed by its immediate neighbors.

\section{Results}
\label{sec:experiments}

In this section, we present the results we obtain by applying \PROD\ using the ODMs of Section \ref{sec:odm} and the people recommenders of Section~\ref{sec:recommenders}.
To foster reproducibility, we publicly release all the code of our experiments at \url{https://github.com/FedericoCinus/PROD-ICWSM2022}.

In the experiments presented next, the parameters are fixed as follows. We generate networks with 400 nodes and $\sim$5500 edges.
In order to avoid disconnected graphs, we set $R_{\max}$ by increasing it in small steps and taking the last value before the graph results disconnected after running \PROD.
In this way, we obtain $R_{\max}=0.4 \cdot |E|$.
However, our results are qualitatively similar with different values of the ratio $R_{\max}/|E|$.
The parameter $T_{max}$ is set to 5000 for BCM and 100 for the epistemological model.
The number of interactions $S$ is set to 2 in order to allow for mixed interactions (both existing edges and recommended ones) at each iteration.
For BCM, the internal parameters $\mu$ and $\epsilon$ are set to $0.2$. For the epistemological model, the parameters $\epsilon$ and $n$ are set to $0.005$ and $15$.
With this choice of parameters, we run the procedure outlined in Section~\ref{sec:framework}, running $K=500$ simulations with each of the given recommenders.
We repeat this process varying the parameters of the random network model, in order to estimate the effect of the recommender over a range of graphs with different initial modularity and homophily; specifically, we test $\eta \in [0.2, 0.8]$ and $\mu \in [0.05, 0.95]$.

\begin{figure*}[t]
 \centering
 \begin{tabular}{cc}
  \hspace{-8mm}\includegraphics[width=.55\linewidth]{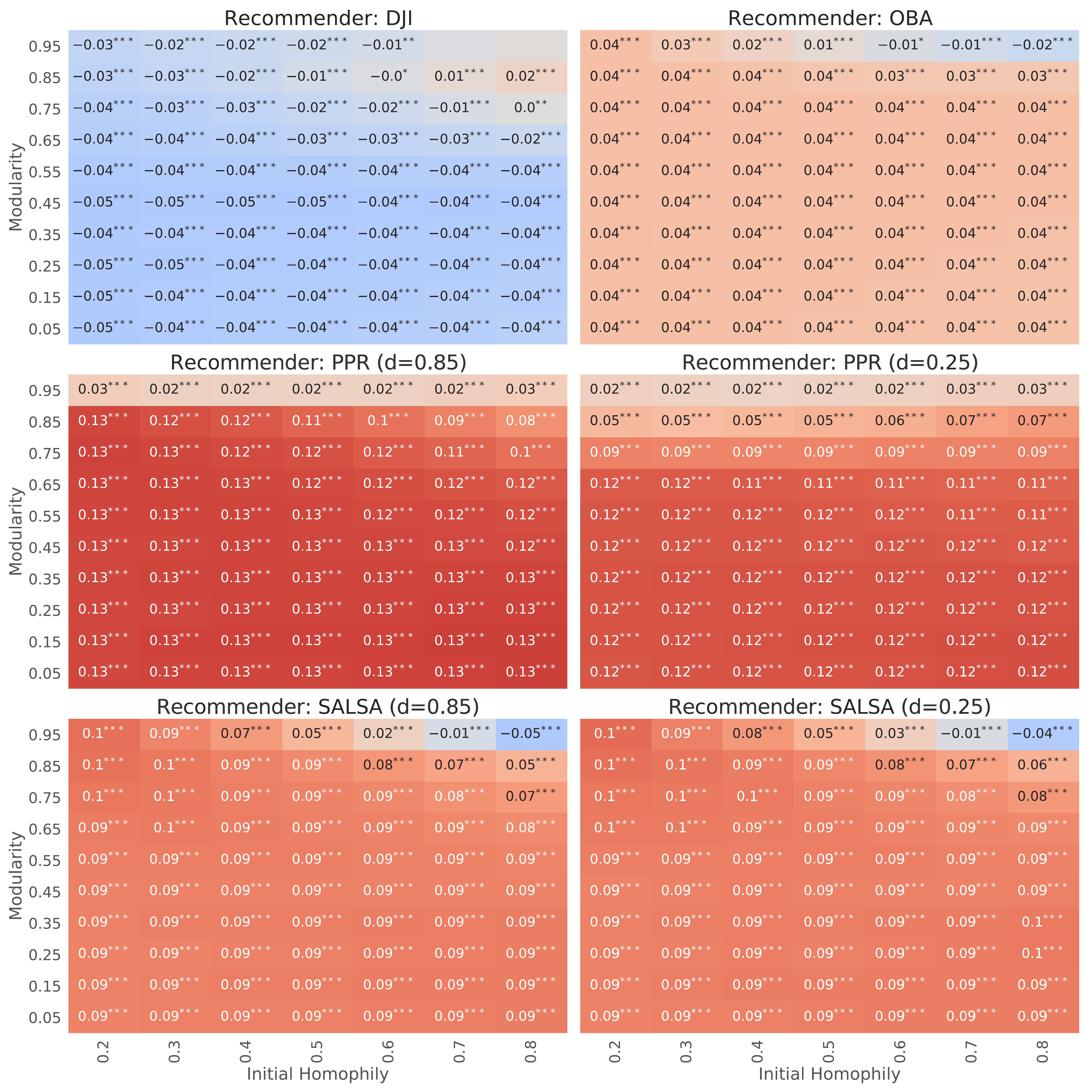}&
 \hspace{-6mm} \includegraphics[width=.55\linewidth]{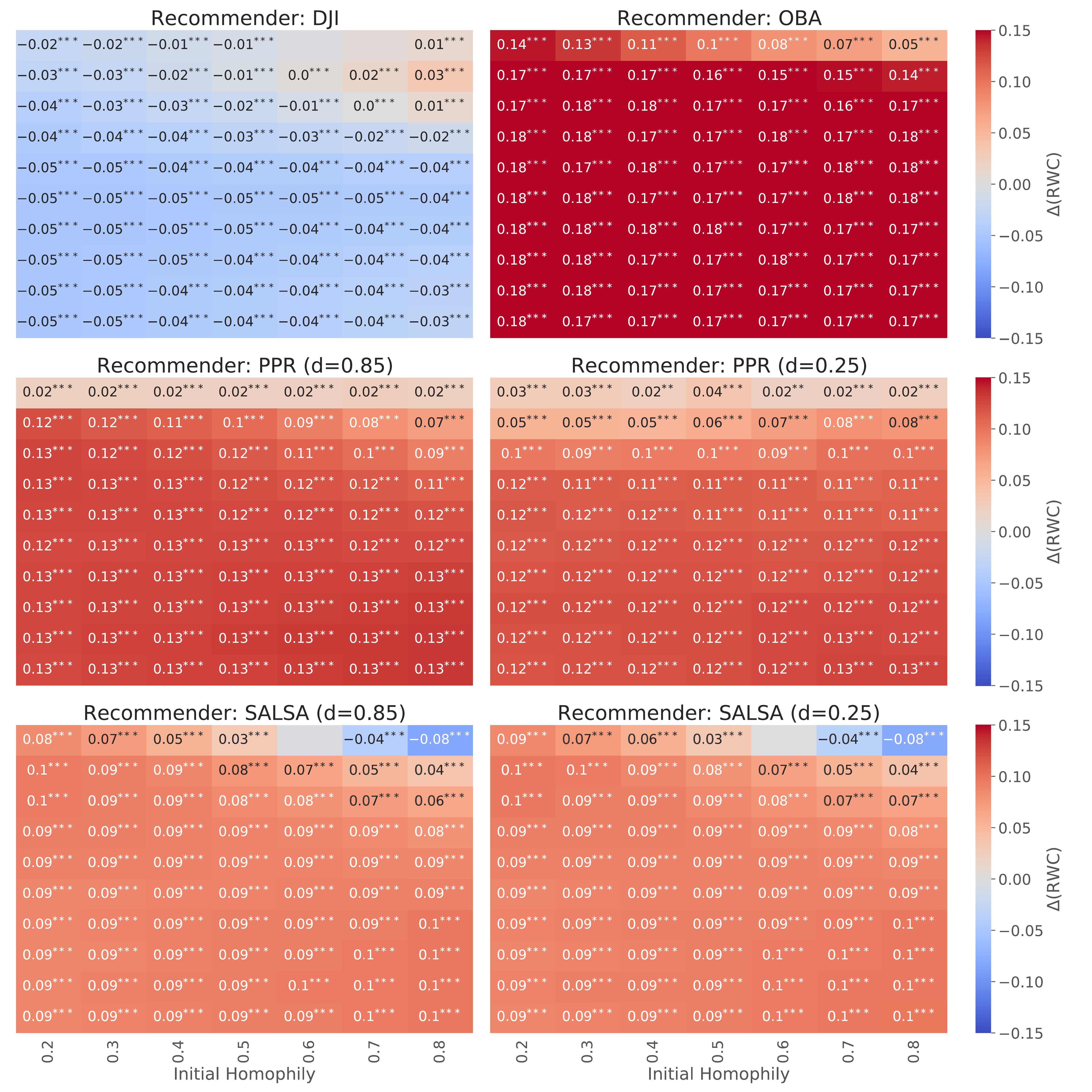}
 \end{tabular}
 \vspace{-2mm}
 \caption{\label{fig:rwc}
 $\Delta$RWC for the epistemological model (left half) and for the BCM model (right half). Each heatmap represents a given recommender and colors represent $\Delta$NCI and $\Delta$RWC
 under different initial conditions (homophily $\eta$ on $x$-axis, modularity $\mu$ on $y$-axis). Numbers are shown when the difference is statistically significant ($p < 0.05$). Asterisks further represent levels of significance (** is $p < 0.01$, *** is $p < 0.001$).}
 \end{figure*}

We present both models' results for the NCI metric in Figure~\ref{fig:nci} and for the RWC metric in Figure~\ref{fig:rwc}.
In each of these figures, the $x$-axis corresponds to the initial homophily $\eta$, while the $y$-axis to the initial modularity $\mu$ of the network.
Therefore, each region of the map corresponds to a certain region of the space of possible initial conditions of the network; examples of these initial conditions are shown in Figure~\ref{fig:random-graphs}.
Since the color represents the difference in our metric between a scenario with a recommender and one without (Equation~\ref{eq:difference}), areas with a more intense shade of red indicate an increase in echo chambers (for NCI) or polarization (for RWC) when a recommender system is used.
In each plot, we mark an entry with a number only if it is significant ($p<0.05$) according to the Kolmogorov-Smirnov test.
Asterisks represent further levels of significance.
Therefore, each of these plots is a \mbox{\emph{fingerprint}} of one recommender system, when considering a specific metric and opinion dynamics model.

In the following, we will analyze first the results we obtained regarding echo-chamber behavior using NCI (Section~\ref{sec:results-nci}), and then regarding polarization using RWC (Section~\ref{sec:results-rwc}).
Then, we explore how our results can change under different assumptions, such as different rewiring rules or susceptibility to recommendations (Section~\ref{sec:generalizability}).
Finally, we show how our framework can be used to assess the impact of intervention policies on recommender systems (Section~\ref{sec:intervention}).

\subsection{Findings on echo chambers}
\label{sec:results-nci}
Let us now focus on the effect of recommenders on echo chambers, as measured by the NCI metric, by looking at Figure~\ref{fig:nci}.
Firstly, we find that all the recommenders significantly increase echo chambers for some initial conditions.
In particular, we always observe an increase when the following two conditions are met:

\begin{enumerate}
  \item[(i)] Homophilic links---i.e., links connecting nodes with the same opinion---are at least half of the initial links of the network (i.e., $\eta>0.5$).
  \item[(ii)] The initial network is not already segregated in polarized communities, but there are a large fraction of inter-community links (i.e., $\mu<0.3$).
\end{enumerate}

For this type of network (highly homophilic, but not well modularized) the effect of people recommenders is always to increase echo chambers, w.r.t. a scenario where no algorithm was introduced.
This finding is consistent across all recommenders and opinion dynamics models.
It is interesting to note, however, that for Personalized PageRank this effect is slightly more intense when the damping factor is lower.
This finding is expected since a lower damping factor corresponds to recommendations being more personalized, while a higher damping factor prioritizes nodes that are central for the whole network.
Therefore, the more recommendations are personalized, the more we find that they can contribute to the rise of echo chambers.

Instead, we observe a different effect in other regions of the initial conditions space.
When the initial graph is not homophilic (i.e., $\eta<0.5$), the effect of the recommender is in general very small, and possibly even in the opposite direction w.r.t. the null model with no recommender.
In other words, the effect of recommenders on homophily is to \emph{amplify} the initial bias present in the network.

A similar---but stronger---pattern can be noticed when the network is highly modularized.
Here, we also observe some differences between different recommenders.
If enough inter-community links are present in the initial network ($\mu > 0.5$), then the effect of DJI is mostly non-significant.
Instead, for PPR and SALSA, we observe a large reduction in the echo-chamber effect when such algorithms are used.
In other words, if the network is not very segregated, recommenders increase homophily and echo chambers; but, if the network is already very divided into communities, some recommenders might have the opposite effect, and ``shuffle together'' different communities.
The algorithms where we observe this effect are those that are able to recommend distant nodes.

Finally, we highlight that these effects hold even for the opinion-biased recommender; however, for this recommender, the overall effect is always of increasing the echo chamber in the network.
This result confirms that such an idealized recommender is useful as a benchmark, showing what happens when recommendations are a result of opinions only, without considering the network structure.

\begin{figure*}[t!]
\centering
\includegraphics[width=\textwidth]{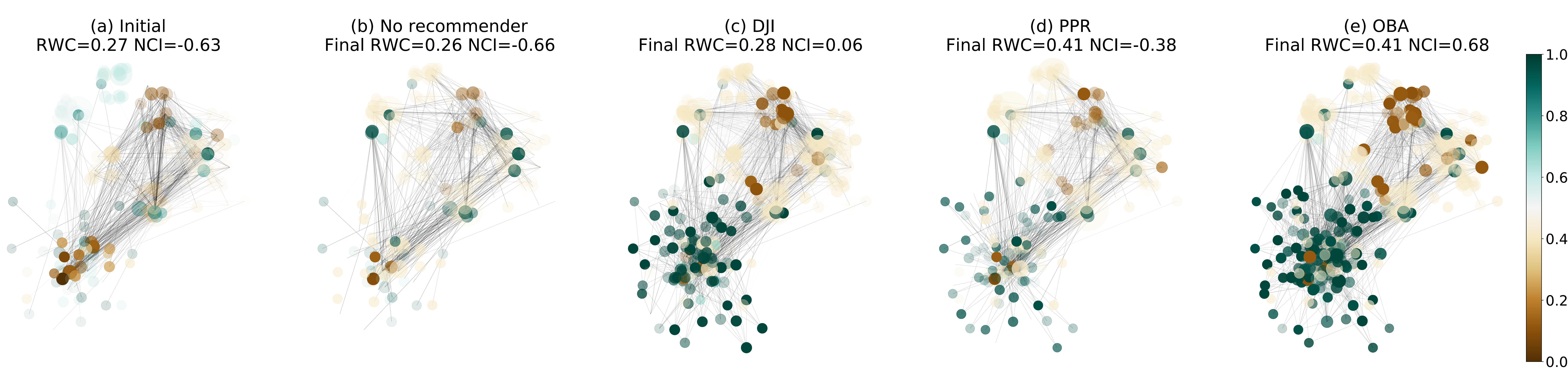}

 \includegraphics[width=\textwidth]{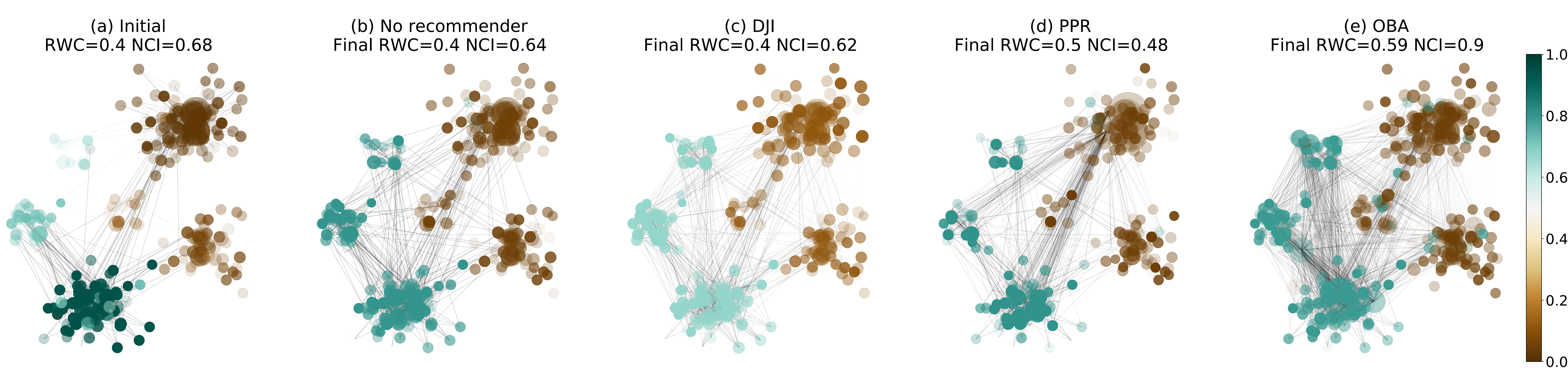}
\vspace{-2mm}
\caption{\label{fig:graph-recommenders} Example of the evolution of networks with two different initial conditions, and three different recommenders. From left to right: (a) initial graph, (b) final graph with no recommender, (c) Directed Jaccard (d) Personalized PageRank, (e) Opinion-Biased. Colors represent the opinion of each node; transparency represents the correlation of their opinions with their neighborhood (i.e., echo chambers). Each community produced by the random network model has an assigned position in the axis. For each node, we draw random coordinates around the correspondent center of the assigned community.
In the first row, the initial configuration is $\eta=0.8,\mu=0.05$ (high homophily, low modularity).
By looking at the difference between (b) and (c,d,e), we observe an increase in echo chamber behavior due to the recommenders.
In the second row with $\eta=0.8,\mu=0.75$, instead, we observe no significant difference.}
\vspace{-2mm}
\end{figure*}

We further investigate our findings in Figure~\ref{fig:graph-recommenders}, by observing a concrete example with two simulated graphs.
Here we highlight which nodes are part of an echo chamber, by setting their transparency as proportional to their contribution in the NCI metric---i.e., how correlated are their opinions to the opinions of their neighbors.
We observe that in high homophily and low modularity setting, echo chambers visibly emerge when a recommender is considered, while they do not in the null model without a recommender.
Instead, when the network is already polarized (i.e., with high initial modularity), the effects of recommenders are negligible.

\subsection{Findings on polarization}
\label{sec:results-rwc}

Now, we turn our attention to the effect of recommenders on polarization, as measured by the RWC metric, by looking at Figure~\ref{fig:rwc}.
First of all, in this case, results are very different between Jaccard Index and random walk-based recommenders, Personalized PageRank and SALSA, for all initial conditions.
In the general case, in fact, while the former type of recommenders obtains a small negative effect on polarization w.r.t. the null model, the latter is increasing it significantly.
We hypothesize that this difference is caused by the information that these two types of recommenders are considering. Since random walk-based recommenders consider higher distances when making recommendations, they prioritize candidates within one's own community, even when such a community is larger than a one-hop neighborhood.
This ends up increasing the separation between the different communities.
Instead, Jaccard Index keeps in consideration only the immediate neighbors of a node; this results in avoiding increasing the segregation between polarized communities in the global structure of the network.
Moreover, since the RWC score is based on how likely is a random walk to cross different communities, it sharply highlights this distinction.
This metric is, therefore, useful to distinguish the higher-order effects of the different recommenders, providing useful information to evaluate the long-term consequences of different choices in recommender algorithms.

Finally, we highlight that for PPR and SALSA the increasing effect holds in all regions of the initial condition space, except for already polarized networks (high~$\mu$, high~$\eta$).
In this case, we observe a saturation phenomenon: the recommender has a smaller effect (for PPR) or even negative (for SALSA) on polarization w.r.t. the null model.
Instead, when DJI starts from such a polarized network, it is the only case when it contributes to further polarization, by amplifying the initial separation.

As in the previous section, all these effects are consistent across the two opinion dynamics models.
Such findings, therefore, are valid both for the case where opinions are indistinguishable and nodes simply influence each other and for the case where nodes are trying to form an opinion on a true fact by exchanging observations.

\newcommand{\ra}[1]{\renewcommand{\arraystretch}{#1}}
\newcommand{\STAB}[1]{\begin{tabular}{@{}c@{}}#1\end{tabular}}
\begin{table*}\centering
\ra{1.3}
\footnotesize
\caption{Comparison between the standard model and two rewiring procedures (opinion-based and degree-based) and two users' susceptibility distributions (uniform and power-law).
EPI and BCM represent the two opinion dynamics models.
Each row represents a pair of values of initial homophily $\eta$ and modularity $\mu$.
We report results for each experiment in terms of our two metrics $\Delta NCI$ and $\Delta RWC$.
Our standard model relies on a uniform rewiring procedure and constant susceptibility for all nodes.
In bold, we highlight cases that differ from the standard model by more than $0.1$.
}
\label{table:generalizability}

\begin{tabular}{@{\extracolsep{\fill}\hspace{\tabcolsep}}
  cccccccccccc
  @{\extracolsep{\fill}\hspace{\tabcolsep}}} \toprule
& & \multicolumn{2}{c}{\textbf{Standard Model}} & \multicolumn{4}{c}{\textbf{Rewiring}}  & \multicolumn{4}{c}{\textbf{Susceptibility}}\\
\cmidrule(lr){3-4} \cmidrule(lr){5-8} \cmidrule(lr){9-12}
& &  &  & \multicolumn{2}{c}{Opinion} & \multicolumn{2}{c}{Degree} & \multicolumn{2}{c}{Uniform} & \multicolumn{2}{c}{Power-law}\\
& $(\eta, \mu)$& EPI & BCM & EPI & BCM & EPI & BCM & EPI & BCM & EPI & BCM  \\
 \midrule
\multirow{4}{*}{\STAB{\rotatebox[origin=c]{90}{\textbf{$\mathbf{\Delta NCI}$}}}}
& $(.2, .05)$& -0.04 & -0.03 & \textbf{0.38} & \textbf{0.64} & -0.06 & -0.05 & -0.03 & -0.03 & -0.03 & -0.03 \\
& $(.2, .95)$ & -0.04 & -0.02 & \textbf{0.23} & \textbf{0.39} & -0.04 & -0.03 & -0.03 & -0.01 & -0.03 & -0.01\\
& $(.8, .05)$ & 0.02 & 0.08 & \textbf{0.42} & \textbf{0.76} & 0.03 & 0.1 & 0.02 & 0.09 & 0.02 & 0.08  \\
& $(.8, .95)$ & -0.01 & 0.00 & 0.04 & 0.08 & -0.01 & 0.00 & -0.01 & 0.00 & -0.01 & 0.00
\\\midrule
\multirow{4}{*}{\STAB{\rotatebox[origin=c]{90}{\textbf{$\mathbf{\Delta RWC}$}}}}
& $(.2, .05)$& 0.13 & 0.13 & 0.16 & \textbf{0.25} & 0.16 & 0.16 & 0.12 & 0.12 & 0.11 & 0.11\\
& $(.2, .95)$ & 0.03 & 0.02 & 0.06 & 0.13 & 0.03 & 0.03 & 0.02 & 0.02 & 0.02 & 0.02\\
& $(.8, .05)$ & 0.13 & 0.13 & 0.16 & \textbf{0.22} & 0.17 & 0.18 & 0.12 & 0.13 & 0.11 & 0.12 \\
& $(.8, .95)$ & 0.03 & 0.02 & 0.11 & \textbf{0.15} & 0.02 & 0.02 & 0.02 & 0.02 & 0.02 & 0.02\\
\bottomrule
\end{tabular}

\end{table*}

\subsection{Generalizability}
\label{sec:generalizability}

In this section, we want to validate the robustness and generalizability of our framework by exploring how changes in the basic assumptions of \PROD\ could impact the results of our analysis. In particular, we evaluate different procedures in terms of rewiring edge selection and individual susceptibility on recommendations.

Following the well-known concept of attention budget~\cite{golder2007rhythms,huberman2008social}, we postulate that each time a new connection is created, an existing one is removed.
In the results we presented so far, we followed the most simple possible choice: select an existing edge with uniform probability, as done in previous work~\citet{sasahara2021social}.
However, it is important to assess to which extent this design choice influences our outcomes.
For this reason, we consider two rewiring policy alternatives:
an \textit{opinion-based} rewiring that picks the edge to remove with a probability proportional to the opinion distance between the two nodes;
and a \textit{degree-based} rewiring, inspired by preferential attachment, where the edge is selected with probability inversely proportional to the degree of the node to unfollow.
We repeat previous experiments with each of these two policies.
For simplicity, we limit this analysis to the PPR recommendation algorithm and to the four extremal configurations of $\eta$ and $\mu$.

We present results in the leftmost columns of Table~\ref{table:generalizability}.
They show quantitatively comparable outcomes between the \textit{standard model}, which employs a random uniform rewiring procedure, and the degree-based policy.
Instead, as one could expect, the rewiring based on the opinion diversity is disruptive, since it greatly amplifies the polarizing effect.
In other words, all our previous findings seem confirmed when the user choices are not directly driven by opinion-based consideration, which of course can explain echo-chamber effects by itself;
however, we have seen how such an assumption is not necessary, since under some configurations recommender systems alone can produce an increase in echo chambers.

Then, we perform experiments challenging our assumption that all nodes have an equal probability $\alpha$ to accept the given recommendation, and observe what happens when each node has an individual \emph{susceptibility} to recommendations.
We test what happens with two different choices to draw this susceptibility, a uniform distribution
where \mbox{$\alpha_u \sim U(0, m)$} and a power-law \mbox{$\alpha_u \sim 1 - m \cdot x^{(m-1)}$}.
In both cases, we set $m$ in order to satisfy $\mathbb{E}[\alpha_u]=\alpha$.
We present these results on the rightmost columns of Table~\ref{table:generalizability}.
By comparing these values with the standard model ones, we find that such a change does not alter significantly our results in all configurations and recommender algorithms.
Similar and consistent results have been achieved with the other recommendation algorithms and in all configurations, thus corroborating the robustness of our findings.

\subsection{Intervention policies}
\label{sec:intervention}

  Finally, in this section, we explore how to use our framework to evaluate the impact of intervention policies in the context of mitigation of echo-chamber and polarization effects.
  We defined three intervention procedures working on top of the recommender system.
  In particular, given an edge from the recommender, we replace it, with probability $\xi$, with another edge selected according to a probability $p$;
  more precisely, we iterate over possible edges in a random order, and for each one evaluate a Bernoullian with probability $p$, until it is successful.
  This probability relies on one of the following strategies.
  \begin{enumerate}
  \item \textit{Random uniform strategy}: $p$ is uniform over the non existing edge set. The original recommendation is therefore modified in order to pursue serendipity.
  \item \textit{Opinion diversity strategy}: $p$ is proportional to the opinion difference of the nodes, i.e. $p = |o_u - o_v|$; here, we try to bridge individuals belonging to different echo chambers.
  \item \textit{Degree-based strategy}: $p$ is higher for nodes with higher degrees, according to a sigmoid: $p = 1 / ( 1 + e^{ -(x - \widehat{d} )})$, where $x$ is the in-degree of the recommended node and $\widehat{d}$ is the average in-degree of the network); the idea is to leverage popular individuals to break echo chambers, as suggested by \citet{elmas2020can}.
  \end{enumerate}

  These three strategies have different degrees of real-world applicability: for instance, a social network user might be more likely to follow a celebrity (i.e. a high-degree node) than a random social media user; following a different-minded individual might be even more unlikely in practice.
  However, for the purpose of this work, we are interested only in analyzing the impact of these idealized policies in mitigating echo-chamber effects.

\begin{figure}[t!]
\centering
\includegraphics[width=\columnwidth]
{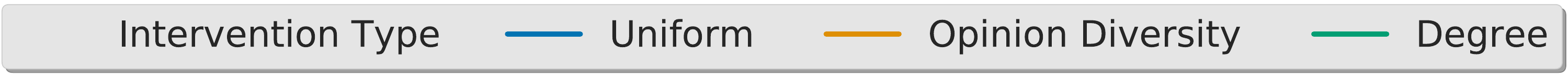}\\
\begin{tabular}{cc}
\hspace{-8mm}
\hspace{-2.75mm}
\includegraphics[scale=0.33]{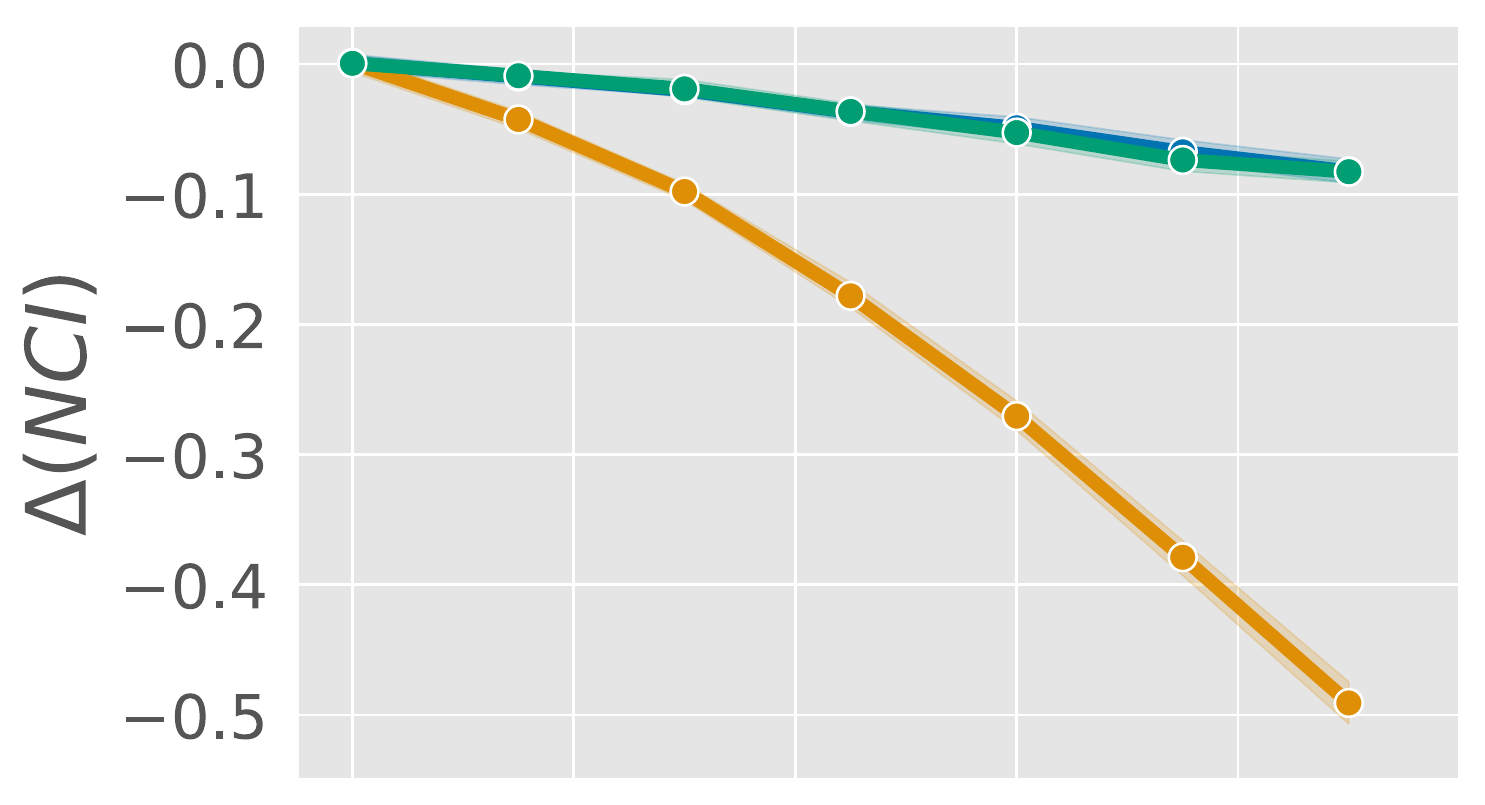}&
\hspace{-8mm}
\hspace{-2.75mm}
\includegraphics[scale=0.33]{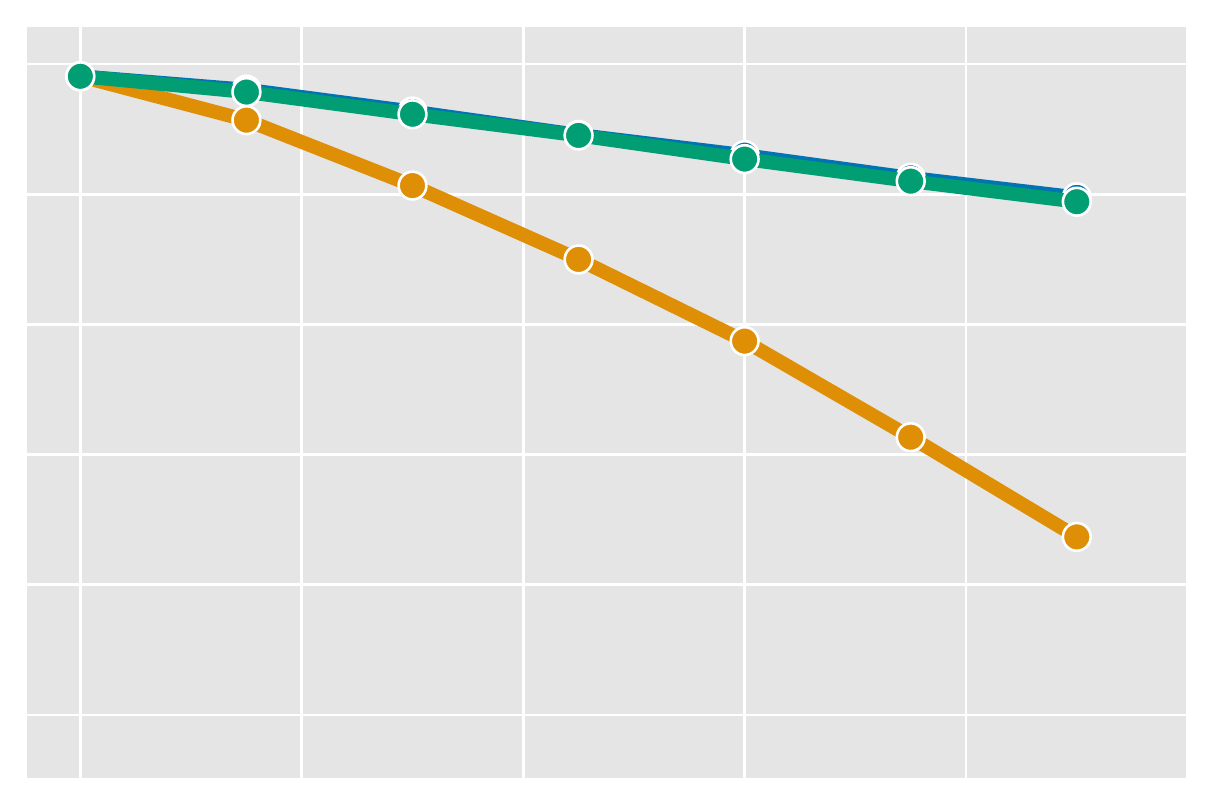}\\
\hspace{-8mm}
\includegraphics[scale=0.33]{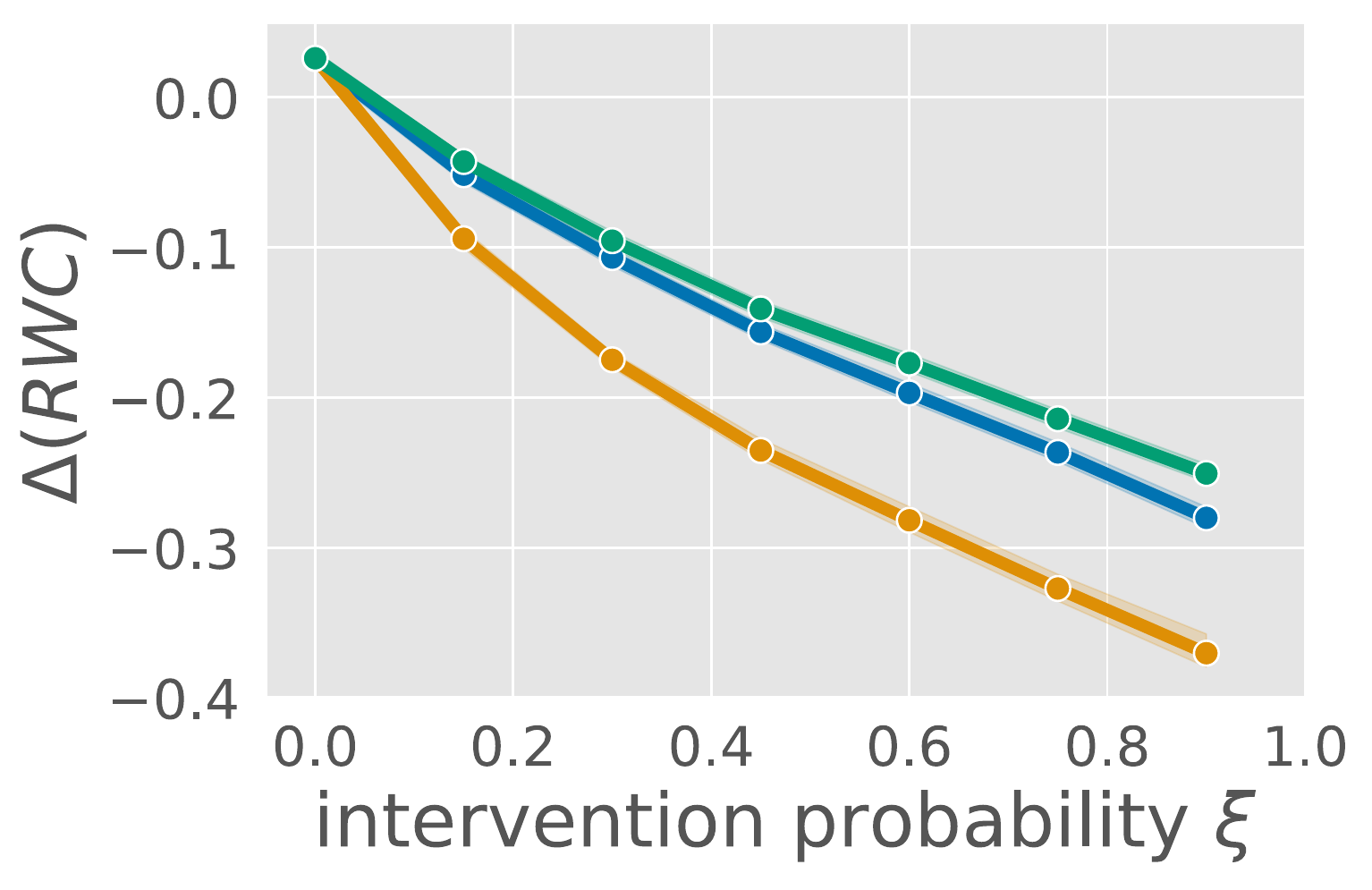}&
\hspace{-8mm}
\includegraphics[scale=0.33]{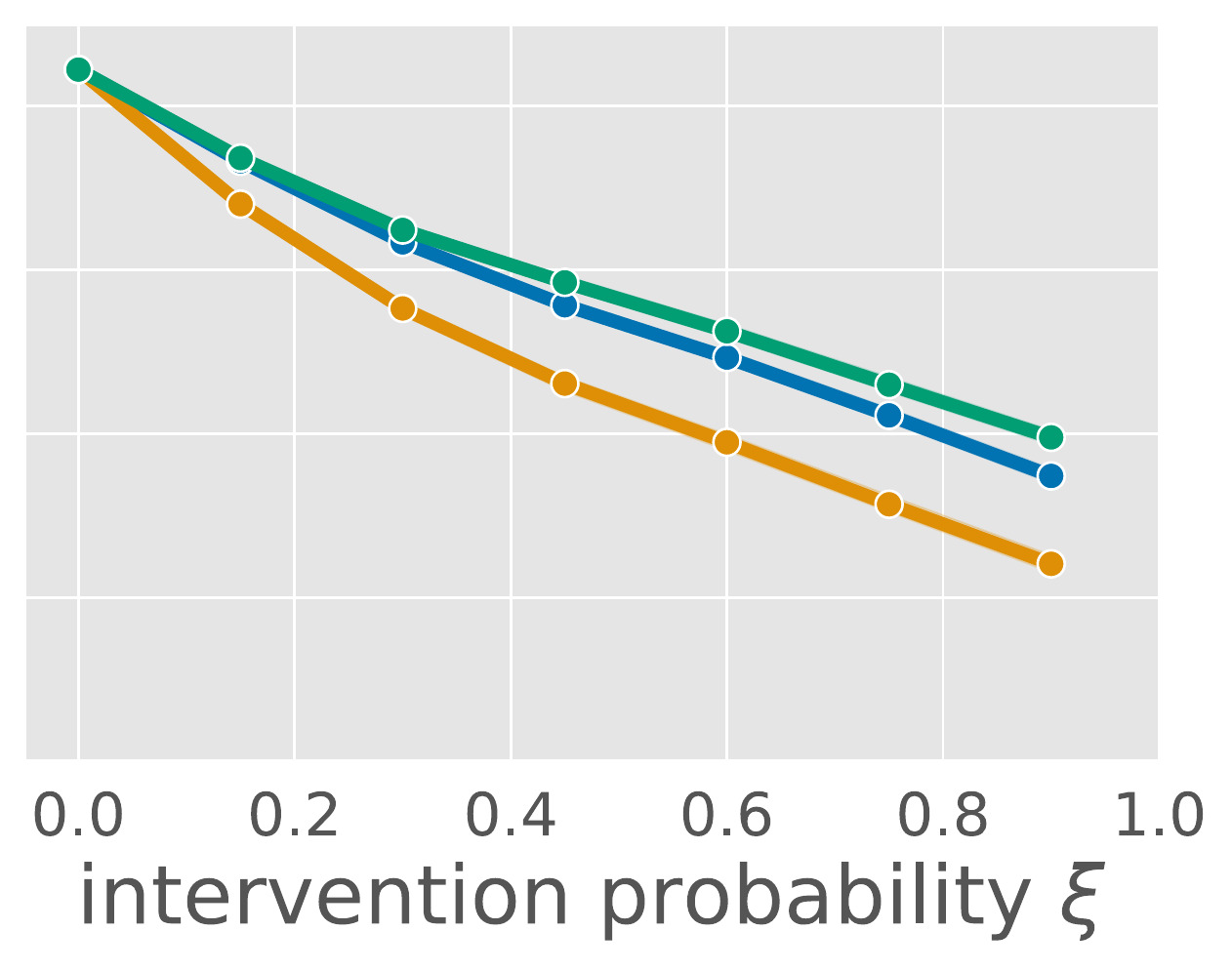}
\end{tabular}
\caption{\label{fig:intervention} $\Delta$NCI (first row) and $\Delta$RWC (second row) as a function of the intervention probability $\xi$, for each intervention procedure and each model: BCM (left half) and Epistemological (right half).}
\end{figure}

  We test these three strategies on top of the personalized PageRank algorithm.
  Results are presented in Figure~\ref{fig:intervention}.
  They show that the in-degree strategy and the uniform strategy have a comparable impact in reducing the echo chamber effect, with a slightly better effect for the latter.
  As expected, the opinion-based strategy results to be more effective because it directly targets the opinions divergence between the users, and this effect is more prominent in terms of the correlation metric (NCI).
  Furthermore, we observe that even low values of intervention probability are already sufficient for reversing the effect on both metrics.

\section{Discussion}
\label{sec:conclusions}
\label{sec:discussion}
In this work, we propose a Monte Carlo simulation procedure that combines opinion dynamics modeling and real-world recommender systems, we quantify the effect of people recommenders in social media platforms on the formation of echo chambers and polarization.
We systematically explore realistic network scenarios, in terms of homophily, modularity, and power-law degree distribution, by extending a previously proposed network model \cite{lancichinettiBenchmarkGraphsTesting2008} by considering opinions on each node.
Thanks to this model, we are able to perform such analysis in a wide range of initial conditions, identifying and exploring two orthogonal dimensions of echo chambers, homophily and modularity.
We consider two complementary opinion-dynamics models, the Bounded Confidence Model~\cite{deffuantMixingBeliefsInteracting2000b} and the epistemological model by~\citet{bala1998learning}.

Our findings are robust across the different opinion dynamics models and people recommenders tested.
We show that the effect of such algorithms is remarkable w.r.t. the null-model and, moreover, prominent when certain initial conditions are met.
The emergence of echo chambers is mostly reinforced by such algorithms when the graph structure is not already modularized, but homophilic connections are prevalent; this finding is consistent for all algorithms, even if with different magnitudes.
The overall polarization shows an increasing trend when the graph is not modularized and tends to a saturation point---with a slightly positive or sometimes negative effect---when the initial conditions are of high homophily and segregation.
In this regard, we are able to highlight a significant difference across recommenders.
We argue that such differences are a result of the tendency of some recommenders to focus on local neighborhoods rather than the global structure. 
Finally, we remark that our framework is able to assess that such findings are in general statistically significant.

More in general, the goal of this work is to define the set of assumptions needed for an increase in echo chamber and polarization to be a direct consequence of recommendation algorithms.
We find that while opinion-biased user behavior can easily explain such effects (Table~\ref{table:generalizability}), in general this assumption is not necessary: the action of recommenders can cause a surge in echo chambers, as long as some initial conditions are met.
Such findings are robust across a wide range of models, algorithms, and design choices.

Nonetheless, we acknowledge that more complex models could be tested: for instance, we assume a static network, where the number of nodes and edges is kept approximately constant over time.
Yet, our experiments with a rewiring rule that resembles preferential attachment yielded very similar results.
We speculate that these kinds of dynamic networks could present similar behavior.

Like other simulation-based studies, our work lacks a direct, quantitative comparison with real-world data, which is obviously difficult to achieve.
In fact, it is impossible in most real cases to observe the counterfactual: 
for a given set of users, simultaneously observing their behavior with and without a recommender system operating.
Our goal is therefore to reproduce phenomena actually observed in real cases, echo-chamber and polarization. 
In order to investigate a causal link between them and recommendation algorithms, it is important, in fact, to have first a formal model of how the two are connected, and to delimit a minimal and realistic set of assumptions for this causal link to exist.
This work is therefore a step in this direction, and as such it opens directions for future research.

Among all, different model extensions regarding multidimensional opinions could be of interest.
From an applicative perspective, our framework could be used as a guideline to design and analyze other existing algorithms in order to assess their interplay with echo chambers.
For instance, one could investigate more complex recommendation algorithms, considering a combination of content-based, graph-based, and interaction-based methods together with recent network embeddings techniques.
Moreover, it can be used to evaluate algorithms able to mitigate echo-chamber effects.
We sketched how such an analysis can work, by comparing the effect of possible intervention strategies.

\bibliographystyle{aaai}
\bibliography{references}

\begin{thebibliography}{}

\bibitem[\protect\citeauthoryear{Bakshy, Messing, and
  Adamic}{2015}]{bakshyExposureIdeologicallyDiverse2015a}
Bakshy, E.; Messing, S.; and Adamic, L.~A.
\newblock 2015.
\newblock Exposure to ideologically diverse news and opinion on {{Facebook}}.
\newblock {\em Science} 348(6239):1130--1132.

\bibitem[\protect\citeauthoryear{Bala and Goyal}{1998}]{bala1998learning}
Bala, V., and Goyal, S.
\newblock 1998.
\newblock Learning from neighbours.
\newblock {\em The review of economic studies} 65(3):595--621.

\bibitem[\protect\citeauthoryear{Barber{\'a} \bgroup et al\mbox.\egroup
  }{2015}]{barberaTweetingLeftRight2015}
Barber{\'a}, P.; Jost, J.~T.; Nagler, J.; Tucker, J.~A.; and Bonneau, R.
\newblock 2015.
\newblock Tweeting {{From Left}} to {{Right}}: {{Is Online Political
  Communication More Than}} an {{Echo Chamber}}?
\newblock {\em Psychological Science} 26(10):1531--1542.

\bibitem[\protect\citeauthoryear{Barbieri, Bonchi, and
  Manco}{2014}]{BarbieriBM14}
Barbieri, N.; Bonchi, F.; and Manco, G.
\newblock 2014.
\newblock Who to follow and why: link prediction with explanations.
\newblock In {\em {KDD} '14},  1266--1275.

\bibitem[\protect\citeauthoryear{Bastos, Mercea, and
  Baronchelli}{2018}]{bastosGeographicEmbeddingOnline2018}
Bastos, M.; Mercea, D.; and Baronchelli, A.
\newblock 2018.
\newblock The geographic embedding of online echo chambers: {{Evidence}} from
  the {{Brexit}} campaign.
\newblock {\em PLOS ONE} 13(11):e0206841.

\bibitem[\protect\citeauthoryear{Cinelli \bgroup et al\mbox.\egroup
  }{2021}]{cinelli2021echo}
Cinelli, M.; Morales, G. D.~F.; Galeazzi, A.; Quattrociocchi, W.; and Starnini,
  M.
\newblock 2021.
\newblock The echo chamber effect on social media.
\newblock {\em Proceedings of the National Academy of Sciences} 118(9).

\bibitem[\protect\citeauthoryear{de Arruda \bgroup et al\mbox.\egroup
  }{2021}]{de2021modeling}
de~Arruda, H.~F.; Cardoso, F.~M.; de~Arruda, G.~F.; Hern{\'a}ndez, A.~R.;
  Costa, L. d.~F.; and Moreno, Y.
\newblock 2021.
\newblock Modeling how social network algorithms can influence opinion
  polarization.
\newblock {\em arXiv preprint arXiv:2102.00099}.

\bibitem[\protect\citeauthoryear{Deffuant \bgroup et al\mbox.\egroup
  }{2000}]{deffuantMixingBeliefsInteracting2000b}
Deffuant, G.; Neau, D.; Amblard, F.; and Weisbuch, G.
\newblock 2000.
\newblock Mixing {{Beliefs Among Interacting Agents}}.
\newblock {\em Advances in Complex Systems} 3:87--98.

\bibitem[\protect\citeauthoryear{Elmas \bgroup et al\mbox.\egroup
  }{2020}]{elmas2020can}
Elmas, T.; Hardi, K.; Overdorf, R.; and Aberer, K.
\newblock 2020.
\newblock Can celebrities burst your bubble?
\newblock {\em MISINFO 2021: Workshop on Misinformation Integrity in Social
  Network}.

\bibitem[\protect\citeauthoryear{Enli}{2017}]{enliTwitterArenaAuthentic2017}
Enli, G.
\newblock 2017.
\newblock Twitter as arena for the authentic outsider: {{Exploring}} the social
  media campaigns of {{Trump}} and {{Clinton}} in the 2016 {{US}} presidential
  election.
\newblock {\em European journal of communication} 32(1):50--61.

\bibitem[\protect\citeauthoryear{Fabbri \bgroup et al\mbox.\egroup
  }{2020}]{FabbriBB020}
Fabbri, F.; Bonchi, F.; Boratto, L.; and Castillo, C.
\newblock 2020.
\newblock The effect of homophily on disparate visibility of minorities in
  people recommender systems.
\newblock In {\em Proceedings of the International AAAI Conference on Web and
  Social Media},  165--175.

\bibitem[\protect\citeauthoryear{Fletcher and
  Kleis~Nielsen}{2017}]{fletcherArePeopleIncidentally2017}
Fletcher, R., and Kleis~Nielsen, R.
\newblock 2017.
\newblock Are people incidentally exposed to news on social media? {{A}}
  comparative analysis.
\newblock {\em New Media \& Society} 20:146144481772417.

\bibitem[\protect\citeauthoryear{Garimella \bgroup et al\mbox.\egroup
  }{2018}]{garimella2018quantifying}
Garimella, K.; Morales, G. D.~F.; Gionis, A.; and Mathioudakis, M.
\newblock 2018.
\newblock Quantifying controversy on social media.
\newblock {\em ACM Transactions on Social Computing} 1(1):1--27.

\bibitem[\protect\citeauthoryear{Garrett}{2009}]{garrett09echo}
Garrett, R.~K.
\newblock 2009.
\newblock Echo chambers online?: {{Politically}} motivated selective exposure
  among {{Internet}} news users.
\newblock {\em JCMC} 14(2).

\bibitem[\protect\citeauthoryear{Gleich}{2015}]{gleich2015pagerank}
Gleich, D.~F.
\newblock 2015.
\newblock Pagerank beyond the web.
\newblock {\em siam REVIEW} 57(3):321--363.

\bibitem[\protect\citeauthoryear{Goel \bgroup et al\mbox.\egroup
  }{2015}]{wtf_2015}
Goel, A.; Gupta, P.; Sirois, J.; Wang, D.; Sharma, A.; and Gurumurthy, S.
\newblock 2015.
\newblock The who-to-follow system at twitter: Strategy, algorithms, and
  revenue impact.
\newblock {\em Interfaces} 45(1):98--107.

\bibitem[\protect\citeauthoryear{Golder, Wilkinson, and
  Huberman}{2007}]{golder2007rhythms}
Golder, S.~A.; Wilkinson, D.~M.; and Huberman, B.~A.
\newblock 2007.
\newblock Rhythms of social interaction: Messaging within a massive online
  network.
\newblock In {\em Communities and technologies 2007}. Springer.
\newblock  41--66.

\bibitem[\protect\citeauthoryear{Gupta \bgroup et al\mbox.\egroup
  }{2013}]{wtf_twitter}
Gupta, P.; Goel, A.; Lin, J.~J.; Sharma, A.; Wang, D.; and Zadeh, R.
\newblock 2013.
\newblock {WTF:} the who to follow service at twitter.
\newblock In {\em 22nd International World Wide Web Conference, {WWW} '13},
  505--514.
\newblock ACM.

\bibitem[\protect\citeauthoryear{Guy}{2018}]{guy2018people}
Guy, I.
\newblock 2018.
\newblock People recommendation on social media.
\newblock In {\em Social information access}. Springer.
\newblock  570--623.

\bibitem[\protect\citeauthoryear{Huberman, Romero, and
  Wu}{2008}]{huberman2008social}
Huberman, B.~A.; Romero, D.~M.; and Wu, F.
\newblock 2008.
\newblock Social networks that matter: Twitter under the microscope.
\newblock {\em arXiv preprint arXiv:0812.1045}.

\bibitem[\protect\citeauthoryear{Khondker}{2011}]{khondkerRoleNewMedia2011}
Khondker, H.~H.
\newblock 2011.
\newblock Role of the new media in the {{Arab Spring}}.
\newblock {\em Globalizations} 8(5):675--679.

\bibitem[\protect\citeauthoryear{Kleinberg}{1999}]{hits}
Kleinberg, J.~M.
\newblock 1999.
\newblock Authoritative sources in a hyperlinked environment.
\newblock {\em J. {ACM}} 46(5):604--632.

\bibitem[\protect\citeauthoryear{Kobourov}{2012}]{kobourov2012spring}
Kobourov, S.~G.
\newblock 2012.
\newblock Spring embedders and force directed graph drawing algorithms.
\newblock {\em arXiv:1201.3011}.

\bibitem[\protect\citeauthoryear{Kumar \bgroup et al\mbox.\egroup
  }{2020}]{kumar2020link}
Kumar, A.; Singh, S.~S.; Singh, K.; and Biswas, B.
\newblock 2020.
\newblock Link prediction techniques, applications, and performance: A survey.
\newblock {\em Physica A: Statistical Mechanics and its Applications}
  553:124289.

\bibitem[\protect\citeauthoryear{Lancichinetti, Fortunato, and
  Radicchi}{2008}]{lancichinettiBenchmarkGraphsTesting2008}
Lancichinetti, A.; Fortunato, S.; and Radicchi, F.
\newblock 2008.
\newblock Benchmark {{Graphs}} for {{Testing Community Detection Algorithms}}.
\newblock {\em Physical Review E} 78(4):046110.

\bibitem[\protect\citeauthoryear{Lempel and Moran}{2001}]{salsa}
Lempel, R., and Moran, S.
\newblock 2001.
\newblock {SALSA:} the stochastic approach for link-structure analysis.
\newblock {\em {ACM} Trans. Inf. Syst.} 19(2):131--160.

\bibitem[\protect\citeauthoryear{Liben-Nowell and
  Kleinberg}{2007}]{liben2007link}
Liben-Nowell, D., and Kleinberg, J.
\newblock 2007.
\newblock The link-prediction problem for social networks.
\newblock {\em Journal of the American society for information science and
  technology} 58(7):1019--1031.

\bibitem[\protect\citeauthoryear{Morales, Monti, and
  Starnini}{2021}]{morales2021no}
Morales, G. D.~F.; Monti, C.; and Starnini, M.
\newblock 2021.
\newblock No echo in the chambers of political interactions on reddit.
\newblock {\em Scientific Reports} 11(1):1--12.

\bibitem[\protect\citeauthoryear{Nikolov \bgroup et al\mbox.\egroup
  }{2015}]{nikolovMeasuringOnlineSocial2015}
Nikolov, D.; Oliveira, D. F.~M.; Flammini, A.; and Menczer, F.
\newblock 2015.
\newblock Measuring online social bubbles.
\newblock {\em PeerJ Computer Science} 1:e38.

\bibitem[\protect\citeauthoryear{Page \bgroup et al\mbox.\egroup
  }{1999}]{page1999pagerank}
Page, L.; Brin, S.; Motwani, R.; and Winograd, T.
\newblock 1999.
\newblock The pagerank citation ranking: Bringing order to the web.
\newblock Technical report, Stanford InfoLab.

\bibitem[\protect\citeauthoryear{Pariser}{2011}]{pariser11filter}
Pariser, E.
\newblock 2011.
\newblock {\em The {{Filter Bubble}}: {{What}} the {{Internet Is Hiding}} from
  {{You}}}.
\newblock {Penguin UK}.

\bibitem[\protect\citeauthoryear{Perra and Rocha}{2019}]{perra2019modelling}
Perra, N., and Rocha, L.~E.
\newblock 2019.
\newblock Modelling opinion dynamics in the age of algorithmic personalisation.
\newblock {\em Scientific reports} 9(1):1--11.

\bibitem[\protect\citeauthoryear{Quattrociocchi, Scala, and
  Sunstein}{2016}]{quattrociocchiEchoChambersFacebook2016}
Quattrociocchi, W.; Scala, A.; and Sunstein, C.~R.
\newblock 2016.
\newblock Echo {{Chambers}} on {{Facebook}}.
\newblock {{SSRN Scholarly Paper}} ID 2795110, {Social Science Research
  Network}, {Rochester, NY}.

\bibitem[\protect\citeauthoryear{Sasahara \bgroup et al\mbox.\egroup
  }{2021}]{sasahara2021social}
Sasahara, K.; Chen, W.; Peng, H.; Ciampaglia, G.~L.; Flammini, A.; and Menczer,
  F.
\newblock 2021.
\newblock Social influence and unfollowing accelerate the emergence of echo
  chambers.
\newblock {\em Journal of Computational Social Science} 4(1):381--402.

\bibitem[\protect\citeauthoryear{S{\^\i}rbu \bgroup et al\mbox.\egroup
  }{2019}]{sirbu2019algorithmic}
S{\^\i}rbu, A.; Pedreschi, D.; Giannotti, F.; and Kert{\'e}sz, J.
\newblock 2019.
\newblock Algorithmic bias amplifies opinion fragmentation and polarization: A
  bounded confidence model.
\newblock {\em PloS one} 14(3):e0213246.

\end{thebibliography}

\end{document}